\documentstyle[12pt,amssym]{article}
\def\N{\mbox{$\Bbb N$}}
\def\ap{a^{\dagger}}
\def\a{a^{\vphantom{\dagger}}}
\def\ddelta{\delta^{\vphantom{\dagger}}}

\title{
\hfill{\normalsize ULB/229/CQ/99/3}\\
\vspace{1cm}
Interpretation and extension of the Green's ansatz for paraparticles}
\author{C. Quesne \thanks{Directeur de recherches FNRS. E-mail:
cquesne@ulb.ac.be}\\
{\small \sl Physique Nucl\'eaire Th\'eorique et Physique
Math\'ematique,  Universit\'e Libre de Bruxelles,} \\
{\small \sl Campus de la
Plaine CP229, Boulevard~du Triomphe, B-1050 Brussels, Belgium}}
\date{ }
\begin{document}
\baselineskip=22pt plus 1pt minus 1pt
\maketitle

\begin{abstract}
The anomalous bilinear commutation relations satisfied by the components of the
Green's ansatz for paraparticles are shown to derive from the
comultiplication of
the paraboson or parafermion algebra. The same provides a generalization of the
ansatz, wherein paraparticles of order $p = \sum_{\alpha=1}^r p_{\alpha}$ are
constructed from $r$ paraparticles of order $p_{\alpha}$, $\alpha=1$, 2,
\ldots,~$r$.
\end{abstract}

\vspace{1.5cm}


\newpage
%
%
In the last few years, there has been increasing interest in generalized
statistics,
different from Bose and Fermi statistics. The main reason is their possible
relevance to the theory of the fractional quantum Hall
effect~\cite{halperin}, to
that of anyon superconductivity~\cite{laughlin}, and to the description of black
hole statistics~\cite{strominger}. There have been various proposals, among
which
we may quote parastatistics~\cite{green, palev94a}, anyon
statistics~\cite{leinaas}, quon statistics~\cite{greenberg}, Haldane fractional
statistics~\cite{haldane}, and also some recent attempts to provide a
unified view
of some of them (see e.g.~\cite{meljanac}).\par
%
%
In the present Letter, we shall be concerned with Green's parabose and parafermi
statistics~\cite{green}, which arose from a remark of Wigner back in
1950~\cite{wigner}, and were among the first consistent examples of generalized
statistics. Green's parastatistics is based upon trilinear commutation
relations for
particle creation and annihilation operators. It is characterized by a discrete
parameter $p \in \N$, called the order of paraquantization and interpolating
between Bose and Fermi statistics. For the paraboson (resp.\ parafermion) case,
only those representations of the symmetric group $S_N$ with at most $p$ rows
(resp.\ columns) do occur, so that at most $p$ parabosons (resp.\ parafermions)
can be in an antisymmetric (resp.\ symmetric) state.\par
%
%
Systems made of a single type of parabosons can be alternatively
described~\cite{chaturvedi, macfarlane} in the framework of the
Calogero-Vasiliev
algebra~\cite{vasiliev} Fock space representation. This algebra also plays
a crucial
role~\cite{poly} in understanding the algebraic properties of the two-particle
Calogero problem~\cite{calogero}. While showing an interesting connection
between
parastatistics and the fast-growing field of integrable models, the equivalence
established in Refs.~\cite{chaturvedi} and~\cite{macfarlane} also provides a
characterization of parabosons in terms of bilinear commutation relations,
hence a
more convenient approach to the Fock space construction.\par
%
%
To deal with the latter problem for systems made of more than one type of
parabosons or for systems of parafermions, one still has to resort to the
Green's
ansatz~\cite{green}, expressing parabosons (resp.\ parafermions) of order~$p$ as
combinations of $p$ anticommuting bosons (resp.\ commuting fermions). A natural
question that arises in connection with such a construction is why the boson or
fermion operators partially commute and partially anticommute. To avoid this
problem, it has been proposed to consider ordinary, i.e., commuting (resp.\
anticommuting), bosons (resp.\ fermions) and to multiply them either by Clifford
matrices~\cite{cusson}, or by Majorana fermions~\cite{macfarlane}. In another
connection, it has been argued that the Green's ansatz construction is a
very natural
method from the Lie superalgebra viewpoint~\cite{palev94b}.\par
%
%
Here we shall adopt another viewpoint, which will at the same time provide an
explanation for the strange behaviour of the Green's ansatz components and
generalize the construction. It is based on the addition of paraboson (resp.\
parafermion) operators or, in mathematical terms, on the comultiplication of the
paraboson (resp.\ parafermion) algebra, which is part of the Hopf algebraic
structure of the latter~\cite{daska}.\par
%
%
Let us recall that the paraboson or parafermion algebra is generated by $n$
pairs of
creation and annihilation operators $\ap_k$, $\a_k$, $k=1$, 2, \ldots,~$n$,
satisfying the trilinear commutation relations~\cite{green}
\begin{eqnarray}
  \left[\a_k, \left[\ap_l, \a_m\right]_{\pm} \right]_- & = & 2 \ddelta_{kl}
\a_m,
         \nonumber\\
  \left[\a_k, \left[\ap_l, \ap_m\right]_{\pm} \right]_- & = & 2
\ddelta_{kl} \ap_m
         \pm 2 \ddelta_{km} \ap_l, \nonumber\\
  \left[a_k, \left[a_l, a_m\right]_{\pm} \right]_- & = & 0,  \label{eq:para-alg}
\end{eqnarray}
where (and in what follows) the upper and lower signs refer to parabosons and
parafermions, respectively, and as usual $[x, y]_{\pm} \equiv xy \pm yx$.\par
%
%
A question that arises is how to build paraboson (resp.\ parafermion) operators
$\ap_k$, $\a_k$, i.e., operators satisfying Eq.~(\ref{eq:para-alg}), out of two
commuting (resp.\ anticommuting) sets of paraboson (resp.\ parafermion)
operators
$\left(\ap_{1k}, \a_{1k}\right)$ and $\left(\ap_{2k}, \a_{2k}\right)$, both
satisfying Eq.~(\ref{eq:para-alg}). It is important to stress here that by
demanding
that $\left[a_{1k}, a_{2l}\right]_{\mp} = \left[\a_{1k}, \ap_{2l}\right]_{\mp} =
0$, we endow the paraboson (resp.\ parafermion) operators with an even (resp.\
odd) character. This seems quite natural having in mind the special case of
bosons
(resp.\ fermions).\par
%
%
If we had a Lie algebra or superalgebra, the answer to the question would be
simple, since we would have $\ap_k = \ap_{1k} + \ap_{2k}$, $\a_k = \a_{1k} +
\a_{2k}$.\footnote{This is valid for instance for boson
operators (see e.g.~\cite{celeghini}).} In mathematical terms, this would
mean that
the coproducts $\Delta\left(\ap_k\right) = \ap_k \otimes I + I \otimes \ap_k$,
$\Delta(\a_k) = \a_k \otimes I + I \otimes \a_k$ would fulfil the same
relations as $\ap_k$, $\a_k$. Here, the symbol $\otimes$ denotes standard tensor
product in the Lie algebraic case, but supertensor product in the Lie superalgebraic
one, i.e., $(x \otimes y) (z \otimes t) = (-1)^{|y| |z|} (xz \otimes yt)$,
where $|y|$ and
$|z|$ are the degrees of $y$ and $z$, respectively (see e.g.\ examples
1.5.7 and 10.1.3
in Ref.~\cite{majid}).\par
%
%
{}For the algebra~(\ref{eq:para-alg}), the problem is more complicated, but was
solved by Daskaloyannis {\em et al.}~\cite{daska} in the paraboson case. In the
parafermion one, these authors assumed that the  operators of the two sets
$\left(\ap_{1k}, \a_{1k}\right)$ and $\left(\ap_{2k}, \a_{2k}\right)$
commute with
one another, so that we shall not follow their solution. Actually, it is
straightforward to see that provided we assume that the two sets of parafermion
operators anticommute, the same comultiplication (and more generally the same
Hopf structure) is valid for parabosons and parafermions.\par
%
%
To define the comultiplication, it is necessary to first extend the paraboson or
parafermion algebra with the operators
\begin{equation}
  K = \exp({\rm i}\pi{\cal N}), \qquad K^{\dagger} = \exp(-{\rm i}\pi{\cal N}),
\end{equation}
where
\begin{equation}
  {\cal N} = \frac{1}{2} \sum_{k=1}^n \left[\ap_k, \a_k\right]_{\pm}.
\end{equation}
As a consequence of Eq.~(\ref{eq:para-alg}), they fulfil the relations
\begin{eqnarray}
  K K^{\dagger} & = & K^{\dagger} K = I, \nonumber \\
  \left[K, \ap_k\right]_+ & = & \left[K, \a_k\right]_+ = \left[K^{\dagger},
         \ap_k\right]_+ = \left[K^{\dagger}, \a_k\right]_+ = 0.  \label{eq:K}
\end{eqnarray}
It is then a simple matter to check that the operators
\begin{eqnarray}
  \Delta(\a_k) & = & \a_k \otimes I + K \otimes \a_k, \qquad \Delta(\ap_k)
= \ap_k
         \otimes I + K^{\dagger} \otimes \ap_k,  \label{eq:coproduct1} \\
  \Delta(K) & = & K \otimes K, \qquad \Delta\left(K^{\dagger}\right) =
K^{\dagger}
         \otimes K^{\dagger},  \label{eq:coproduct2}
\end{eqnarray}
satisfy both the trilinear relations~(\ref{eq:para-alg}) and the additional
relations~(\ref{eq:K}) if
\begin{equation}
  [\a_k \otimes I, I \otimes \a_l]_{\mp} = \left[\a_k \otimes I, I \otimes
  \ap_l\right]_{\mp} = 0,
\end{equation}
which is consistent with the even (resp.\ odd) character of paraboson (resp.\
parafermion) operators.
\par
%
%
Equation~(\ref{eq:para-alg}) only determines the parabosonic or parafermionic
nature of the creation and annihilation operators. To fix the order of
paraquantization $p$, one has to impose the additional condition
\begin{equation}
  \a_k \ap_l |0\rangle = \ddelta_{kl}\, p\, |0\rangle,
\end{equation}
where $|0\rangle$ is the parabosonic or parfermionic vacuum state, i.e.,
\begin{equation}
  \a_k |0\rangle = 0.
\end{equation}
\par
%
%
Let us assume that we start from two sets of paraboson or parfermion operators
$\left(\ap_{1k}, \a_{1k}\right)$, $\left(\ap_{2k}, \a_{2k}\right)$ with
fixed orders
of paraquantization $p_1$ and $p_2$, respectively, or in other words that
\begin{equation}
  \a_{\alpha k} \ap_{\alpha l} |0\rangle_{\alpha} = \ddelta_{kl}\, p_{\alpha}\,
  |0\rangle_{\alpha}, \qquad \a_{\alpha k} |0\rangle_{\alpha} = 0, \qquad
\alpha = 1,
  2.
\end{equation}
What can we say from their combination? By using Eq.~(\ref{eq:coproduct1}), we
obtain that $|0\rangle_1 |0\rangle_2 = |0\rangle \otimes |0\rangle$ is the
vacuum
state of the combined operators,
\begin{equation}
  \Delta(\a_k) (|0\rangle \otimes |0\rangle) = 0,  \label{eq:vacuum}
\end{equation}
and moreover that
\begin{equation}
  \Delta(\a_k) \Delta(\ap_l) (|0\rangle \otimes |0\rangle) = \ddelta_{kl} (p_1 +
  p_2) (|0\rangle \otimes |0\rangle).  \label{eq:order}
\end{equation}
This shows that the combined operators are parabosons or parafermions of order
$p_1+p_2$.\par
%
%
The results obtained so far for the addition of two types of paraboson or
parafermion operators can be easily extended to that of $r$ types of such
operators.
This implies iterating the comultiplication defined in
Eqs.~(\ref{eq:coproduct1})
and~(\ref{eq:coproduct2}). Such a process is made possible by an important
property of the comultiplication valid for coalgebras (and more generally
for Hopf
algebras), called coassociativity, according to which the order wherein the
addition
of three types of operators is performed does not matter:
\begin{equation}
  ({\rm id} \otimes \Delta) \Delta(X) = (\Delta \otimes {\rm id}) \Delta(X).
  \label{eq:coassoc}
\end{equation}
It is straightforward to check that Eqs.~(\ref{eq:coproduct1})
and~(\ref{eq:coproduct2}) indeed satisfy condition~(\ref{eq:coassoc}).\par
%
%
Hence, we may recursively define $\Delta^{(r-1)} \equiv \left(\Delta \otimes
I^{(r-2)}\right) \Delta^{(r-2)}$, $\Delta^{(1)} \equiv \Delta$, where
$I^{(r-2)} \equiv
I \otimes I \otimes \cdots \otimes I$ ($r-2$ times), and construct the operators
\begin{equation}
  \Delta^{(r-1)}(\a_k) = \sum_{\alpha=1}^r K^{(\alpha-1)} \otimes \a_k \otimes
  I^{(r-\alpha)}, \qquad    \Delta^{(r-1)}\left(\ap_k\right) = \sum_{\alpha=1}^r
  \left(K^{\dagger}\right)^{(\alpha-1)} \otimes \ap_k \otimes I^{(r-\alpha)},
  \label{eq:iter-copro}
\end{equation}
and
\begin{equation}
  \Delta^{(r-1)}(K) = K^{(r)}, \qquad \Delta^{(r-1)}\left(K^{\dagger}\right) =
  \left(K^{\dagger}\right)^{(r)},
\end{equation}
satisfying Eqs.~(\ref{eq:para-alg}) and~(\ref{eq:K}), with $K^{(\alpha)}
\equiv K
\otimes K \otimes \cdots \otimes K$, $\left(K^{\dagger}\right)^{(\alpha)} \equiv
\left(K^{\dagger}\right) \otimes \left(K^{\dagger}\right) \otimes \cdots \otimes
\left(K^{\dagger}\right)$ ($\alpha$ times).\par
%
%
Equations~(\ref{eq:vacuum}) and~(\ref{eq:order}) can also be extended to
\begin{equation}
  \Delta^{(r-1)}(\a_k)\, |0\rangle^{(r)} = 0,
\end{equation}
\begin{equation}
  \Delta^{(r-1)}(\a_k) \Delta^{(r-1)}\left(\ap_l\right)\, |0\rangle^{(r)} =
  \ddelta_{kl} \left(\sum_{\alpha=1}^r p_{\alpha}\right) |0\rangle^{(r)},
\end{equation}
where $|0\rangle^{(r)} \equiv |0\rangle \otimes |0\rangle \otimes \cdots \otimes
|0\rangle$ ($r$ times), and the operators $\ap_k$, $\a_k$ belonging to the
$\alpha$th subspace of the tensor product (or the $\alpha$th set) are
assumed to be
of order $p_{\alpha}$.\par
%
%
Let us now consider the $r$ components of the iterated coproducts of $\a_k$ and
$\ap_k$, defined in Eq.~(\ref{eq:iter-copro}),
\begin{equation}
  a_k^{(\alpha)} = K^{(\alpha-1)} \otimes \a_k \otimes I^{(r-\alpha)}, \qquad
  a_k^{(\alpha)\dagger} = \left(K^{\dagger}\right)^{(\alpha-1)} \otimes \ap_k
  \otimes I^{(r-\alpha)}, \qquad \alpha = 1, 2, \ldots, r.  \label{eq:iter-comp}
\end{equation}
\par
%
%
In the special case where the operators $\ap_k$, $\a_k$ are boson or fermion
operators, i.e., $p_{\alpha} = 1$, $\alpha = 1$, 2, \ldots,~$r$, it results from
Eq.~(\ref{eq:iter-comp}) and from the properties of the operators $\a_k$,
$\ap_k$,
$K$, $K^{\dagger}$ that
\begin{eqnarray}
  \left[a_k^{(\alpha)}, a_l^{(\alpha)\dagger}\right]_{\mp} & = &
\ddelta_{kl}\, I^{(r)},
         \qquad \left[a_k^{(\alpha)}, a_l^{(\alpha)}\right]_{\mp} = 0,
         \label{eq:green1} \\
  \left[a_k^{(\alpha)}, a_l^{(\beta)\dagger}\right]_{\pm} & = &
\left[a_k^{(\alpha)},
         a_l^{(\beta)}\right]_{\pm} = 0 \qquad (\alpha \ne \beta),
\label{eq:green2}
\end{eqnarray}
and
\begin{equation}
  a_k^{(\alpha)} |0\rangle^{(r)} = 0. \label{eq:green3}
\end{equation}
Hence, the components $a_k^{(\alpha)}$, $a_k^{(\alpha)\dagger}$ of the iterated
coproducts satisfy the same anomalous bilinear commutation relations as the
components of the Green's ansatz: in the paraboson (resp.\ parafermion) case,
$a_k^{(\alpha)\dagger}$, $a_k^{(\alpha)}$ are boson (resp.\ fermion)
creation and
annihilation operators, but different sets of operators anticommute (resp.\
commute) among themselves. This shows that the iterated coproducts
$\Delta^{(p-1)}(\a_k)$, $\Delta^{(p-1)}\left(\ap_k\right)$, where we now assume
$r=p$, are but a realization of the Green's ansatz for paraparticles.\par
%
%
{}Furthermore, Eq.~(\ref{eq:iter-comp}) allows us to derive a more general
result.
In the case where the $p_{\alpha}$'s are arbitrary, we indeed find that
Eq.~(\ref{eq:green1}) has to be replaced by
\begin{eqnarray}
  \left[a_k^{(\alpha)}, \left[a_l^{(\alpha\dagger},
a_m^{(\alpha)}\right]_{\pm} \right]
          & = & 2 \ddelta_{kl} a_m^{(\alpha)}, \nonumber\\
  \left[a_k^{(\alpha)}, \left[a_l^{(\alpha\dagger},
a_m^{(\alpha\dagger)}\right]_{\pm}
          \right] & = & 2 \ddelta_{kl} a_m^{(\alpha)\dagger} \pm 2 \ddelta_{km}
          a_l^{(\alpha)\dagger}, \nonumber\\
  \left[a_k^{(\alpha)}, \left[a_l^{(\alpha}, a_m^{(\alpha)}\right]_{\pm} \right]
          & = & 0,
\end{eqnarray}
while Eq.~(\ref{eq:green2}) remains valid, and Eq.~(\ref{eq:green3}) is
supplemented
by
\begin{equation}
  a_k^{(\alpha)} a_l^{(\alpha)\dagger} |0\rangle^{(r)} = \ddelta_{kl}\,
p_{\alpha}
  |0\rangle^{(r)}.
\end{equation}
Paraboson (resp.\ parafermion) operators of order $p = \sum_{\alpha=1}^r
p_{\alpha}$ can therefore be constructed from $r$ anticommuting (resp.\
commuting) sets of paraboson (resp.\ parafermion) operators of order $p_1$,
$p_2$,
\ldots,~$p_r$, respectively.\par
%
%
In conclusion, we did prove that the Green's ansatz for paraparticles has
its origin
in the Hopf algebraic structure of the paraboson or parafermion algebra. In
such a
context, the fact that this algebra is not a Lie algebra, but is defined in
terms of
trilinear relations, plays a crucial role. In addition, we showed that the
ansatz is
but a special case of a more general one, where bosons (resp.\ fermions) are
replaced by parabosons (resp.\ parafermions) of order $p_{\alpha}$, $\alpha
= 1$, 2,
\ldots,~$r$, with $p = \sum_{\alpha=1}^r p_{\alpha}$.\par
%
%


\newpage
\begin{thebibliography}{99}

\bibitem{halperin} B.I.\ Halperin, Phys.\ Rev.\ Lett.\ 52 (1984) 1583, 2390(E).

\bibitem{laughlin} R.B.\ Laughlin, Phys.\ Rev.\ Lett.\ 60 (1988) 2677.

\bibitem{strominger} A.\ Strominger, Phys.\ Rev.\ Lett.\ 71 (1993) 3397.

\bibitem{green} H.S.\ Green, Phys.\ Rev.\ 90 (1953) 270;\\
O.W.\ Greenberg and A.M.L.\ Messiah, Phys.\ Rev.\ B 138 (1965) 1155; J.\ Math.\
Phys.\ 6 (1965) 500;\\
Y.\ Ohnuki and S.\ Kamefuchi, Quantum field theory and parastatistics (Springer,
Berlin, 1982).

\bibitem{palev94a} T.D.\ Palev and N.I.\ Stoilova, J.\ Phys.\ A 27 (1994)
977, 7387;
J.\ Math.\ Phys.\ 38 (1997) 2506.

\bibitem{leinaas} J.M.\ Leinaas and J.\ Myrheim, Nuovo Cimento B 37 (1977) 1;\\
F.\ Wilczek, Phys.\ Rev.\ Lett.\ 49 (1982) 957.

\bibitem{greenberg} O.W.\ Greenberg, Phys.\ Rev.\ Lett.\ 64 (1990) 705; Phys.\
Rev.\ D 43 (1991) 4111.

\bibitem{haldane} F.D.M.\ Haldane, Phys.\ Rev.\ Lett.\ 67 (1991) 937.

\bibitem{meljanac} S.\ Meljanac and M.\ Milekovi\'c, Int.\ J.\ Mod.\ Phys.\ A 11
(1996) 1391;\\
S.\ Meljanac, M.\ Stoji\'c and M.\ Milekovi\'c, Mod.\ Phys.\ Lett.\ A 13
(1998) 995;\\
S.\ Meljanac, M.\ Milekovi\'c and M.\ Stoji\'c, J.\ Phys.\ A 32 (1999) 1115.

\bibitem{wigner} E.P.\ Wigner, Phys.\ Rev.\ 77 (1950) 711.

\bibitem{chaturvedi} S.\ Chaturvedi and V.\ Srinivasan, Phys.\ Rev.\ A 44 (1991)
8024.

\bibitem{macfarlane} A.J.\ Macfarlane, J.\ Math.\ Phys.\ 35 (1994) 1054.

\bibitem{vasiliev} M.A.\ Vasiliev, Int.\ J.\ Mod.\ Phys.\ A 6 (1991) 1115.

\bibitem{poly} A.P.\ Polychronakos, Phys.\ Rev.\ Lett.\ 69 (1992) 703;\\
L.\ Brink, T.H.\ Hansson and M.A.\ Vasiliev, Phys.\ Lett.\ B 286 (1992) 109.

\bibitem{calogero} F.\ Calogero, J.\ Math.\ Phys.\ 10 (1969) 2191, 2197; 12 (1971)
419.

\bibitem{cusson} R.Y.\ Cusson, Ann.\ Phys.\ (NY) 55 (1969) 22;\\
O.\ W.\ Greenberg and K.\ I.\ Macrae, Nucl.\ Phys.\ B 219 (1983) 538.

\bibitem{palev94b} T.D.\ Palev, J.\ Phys.\ A 27 (1994) 7373.

\bibitem{daska} C.\ Daskaloyannis, K.\ Kanakoglou and I.\ Tsohantjis, Hopf
algebraic
structure of the parabosonic and parafermionic algebras and paraparticle
generalization of the Jordan Schwinger map, preprint math-ph/9902005 (1999).

\bibitem{celeghini} E.\ Celeghini, R.\ Giachetti, E.\ Sorace and M.\
Tarlini, J.\
Math.\ Phys.\ 32 (1991) 1155.

\bibitem{majid} S.\ Majid, Foundations of Quantum Group Theory (Cambridge
University, Cambridge, 1995).

\end{thebibliography}
\end{document}